\documentstyle[aps,prb,graphicx]{revtex}
\newif\ifbuidln
\buidlntrue

\begin{document}
\title{Instabilities of interacting electrons on the triangular lattice}
\author{Carsten Honerkamp}
\address{Department of Physics, Massachusetts Institute of Technology, Cambridge
MA 02139, USA } 
\date{April 24, 2003}
\maketitle
\begin{abstract} 
Motivated by the recent finding of superconductivity in layered CoO$_2$ compounds, we investigate superconducting and magnetic instabilities of interacting 
electrons on the two-dimensional triangular lattice. Using a one-loop renormalization group scheme for weak to moderate coupling strengths, we find that for purely local interactions $U>0$ and small Fermi surfaces the renormalization group flow remains bounded down to very low scales and no superconducting or other instabilities can be detected. Antiferromagnetic exchange interactions $J$ generate a wide density region with a $d_{x^2-y^2}+id_{xy}$-wave superconducting instability similar to recent proposals for the strongly correlated $t$-$J$ model.
For larger Fermi surface volumes the interactions flow to strong coupling also for purely local interactions $U>0$. We find a singlet pairing instability in the vicinity of strong magnetic ordering tendencies at three wavevectors for the van Hove filling.  
\end{abstract}
\section{Introduction}
The triangular lattice has gained interest in modern condensed matter physics due to several reasons. First the triangular lattice continues to be important in the study of frustration effects in spin models, and is a promising candidate for the occurrence of spin-liquid phases\cite{andersonfazekas,misguich}. 
Then  there are organic superconductors like the $\kappa$-(BEDT-TTF) materials that may be described by an anisotropic triangular lattice and whose phase diagrams show some resemblance to those of the high-$T_c$ cuprates\cite{mckenzie}.  
Very recently superconductivity was discovered at critical temperatures of 5K\cite{takada} in Na$_{0.35}$CoO$_2\cdot 1.3$H$_2$O.
% and indications for superconductivity were observed at temperatures as high as 31K\cite{wen} in An$_x$CoO$_{2+\delta}$ (An$=$ Na, K). 
In this compound edge-sharing octahedra of CoO$_2$ form two-dimensional layered  triangular networks. A one-band Hubbard model on the two-dimensional triangular lattice may be the simplest approach to describe the motion of the charge carriers in the layers. 

In the cobalt oxides there are signs of electronic correlations like weak ferromagnetism and high temperature Curie behavior of the susceptibility\cite{tsukuda,motohashi}. Therefore it is natural to ask whether the superconductivity in these materials arises due to electron-electron interactions. The experimental findings have immediately spurred several theoretical suggestions\cite{baskaran,kumar,wang,ogata} that RVB (resonating valence bond) superconductivity\cite{anderson} might be realized in these materials. The RVB theories in the framework of the $t$-$J$-model consider the case of strong on-site repulsion on the Co sites. 
An alternative approach is to start from weak coupling. A very flexible way to investigate superconductivity arising from weak to moderate electron-electron interactions is to use renormalization group (RG) techniques. Modern functional RG studies have proved useful in the analysis of the Hubbard model on the  two-dimensional square lattice\cite{zanchi,halboth,honerkamp,honerkampfm} and give results that are in agreement with experiments and other theoretical approaches. In particular, on the square lattice for densities close to one electron per site, various strong coupling approaches\cite{sorella,gauge} and weak coupling RG calculations\cite{zanchi,halboth,honerkamp,honerkampfm} agree on the symmetry of the superconducting order parameter.

The Hubbard model on the triangular lattice has been analyzed using RG techniques by Tsai and Marston\cite{tsai}. These authors were mainly concerned with the half-filled case and could not find any relevant superconducting tendencies for the isotropic triangular lattice. Vojta\cite{vojta} and Dagotto studied superconducting  instabilities due to paramagnon-exchange and found that the $d_{xy}$ channel dominates for most densities less than 1 electron/site. In addition they obtained an odd-frequency $s$-wave pairing as dominant channel close to half band filling.  The fluctuation-exchange (FLEX) method was also applied to search for superconductivity on the anisotropic triangular lattice\cite{kino}.  
In this paper we study two cases: 

a) We analyze the parameter case that is, based on LDA (local density approximation) calculations for NaCo$_2$O$_4$\cite{singh}, currently believed to be relevant for the CoO$_2$ compound where superconductivity was found. 
NaCo$_2$O$_4$ has 1.5 electrons per Co site in the $A_{1g}$-$t_{2g}$ band
close to the Fermi energy. For more details see the papers of Singh\cite{singh} and Wang et al.\cite{wang}. 
The superconducting samples (Na$_{0.35}$CoO$_2\cdot 1.3$H$_2$O) should have 0.15 electrons less per Co atom than in the NaCo$_2$O$_4$-layer, i.e. $\langle n \rangle \approx 1.35$/Co site. If we want to describe the LDA band structure with a tight binding dispersion, the hopping amplitude $t$ should be chosen negative. Then the Fermi surface (FS) is hole-like and closed around the origin, in agreement with photoemission data on NaCo$_2$O$_4$\cite{valla}.
In our model calculation we choose $t>0$. The case $t<0$ is obtained by interchanging particles and holes. Then the superconducting samples correspond to $\langle n \rangle \approx 0.65$/site. We also study the half-filled band, $\langle n \rangle =1$/site, as this is interesting to make contact with other work. Furthermore we discuss the effect of exchange interactions between nearest neighbors.

b) The triangular lattice for $t>0$ at $3/4$ band filling ($\langle n \rangle =1.5$/site) has a van Hove singularity in the density of states at the Fermi level, and the hexagon-shaped Fermi surface is nested. Thus we expect some kind of infrared instability of the Landau-Fermi liquid state. $\vec{k}$-space resolved $N$-patch RG may be the appropriate method for an unbiased description of the flow to strong coupling.

The paper is organized as follows. In Sec. II we describe the $N$-patch renormalization group scheme and its implementation, in Sec. III we discuss case a) with small to medium band filling, and in Sec. IV we treat b), i.e. the density region near the van Hove filling. We conclude with some remarks on a possible relation to the cobalt oxides in Sec. V.   

\section{The model and the method}
\begin{figure}
\begin{center} 
\ifbuidln\includegraphics[width=.69\textwidth]{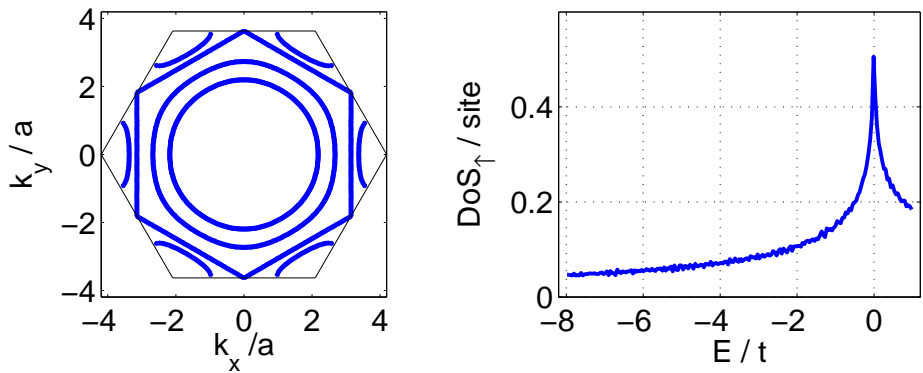}\fi
\end{center} 
\caption{Left: Brillouin zone and lines of constant band energy for the triangular lattice. The volumes encircled by these lines correspond to density $n=0.65$, $n=1$, $n=1.5$ and $n=1.8$ particles per site. Right: Density of states per spin and site. The chemical potential is chosen corresponding to $n=1.5$ and the Fermi energy is at the van Hove density.}
\label{bzdos}
\end{figure}

The Hamiltonian of the Hubbard model on the 2D triangular lattice is
\[ H= - t \sum_{\langle ij \rangle } \left[ c_{i,s}^\dagger c_{j,s} + c_{j,s}^\dagger c_{i,s}\right] + U \, \sum_i  n_{i,\uparrow} n_{i \downarrow} \, . \]
Here $t$ is the amplitude for hopping between nearest neighbors $\langle ij \rangle$ on the triangular lattice and $U>0$ is the on-site repulsion. We choose the reciprocal lattice to be spanned by the vectors $(2 \pi,\pm 4\pi/\sqrt{3})$ where we have set the lattice spacing to unity. Then the first Brillouin zone (BZ) is a regular hexagon (see Fig. \ref{bzdos}). 
The non-interacting  kinetic energy reads 
\[ \epsilon (\vec{k} ) = -2 t \left[ \cos k_x + 2 \cos \frac{k_x}{2} \cos \frac{\sqrt{3}}{2} k_y \right] - \mu \, , \]
where $\mu$ is the chemical potential regulating the particle density. The total bandwidth is $9t$. For the cobalt oxides $t<0$ according to LDA calculations\cite{singh}. In this paper we will take $t>0$. In order to treat the case $t<0$ we have to interchange particles and holes.

Next we briefly 
describe the RG formalism that is used to obtain the effective 
interactions at low temperatures. 
We apply the so-called temperature-flow RG scheme introduced
recently~\cite{honerkampfm} in a $N$-patch implementation that covers
the full Fermi surface.  Similar to the approaches in
Refs.~\onlinecite{zanchi,halboth,honerkamp} the $T$-flow scheme is
derived from an exact RG equation. However, no low energy cutoff is introduced. 
Instead of integrating out modes we use the temperature $T$ itself  as flow
parameter and follow the evolution of the fermionic vertex 
functions as $T$ is varied. 
As was argued recently\cite{honerkampfm} the procedure allows 
an unbiased comparison between AF and FM tendencies. In contrast with that,
 RG schemes with a flowing IR cutoff artificially suppress particle-hole 
excitations with small wave vectors, e.g. long wavelength density fluctuations.  
For a derivation and discussion of the $T$-flow RG scheme 
the reader is referred to Ref.~\onlinecite{honerkampfm}. 

The RG formalism\cite{sh} produces a hierarchy of differential equations for the
one-particle irreducible $n$-point vertex functions, $\Gamma_T^{(n)}$,
as functions of the temperature. Integration of this system of
equations gives the temperature-flow. As initial condition we assume
that at a high temperature $T_0$, the single-particle Green's function
of the system is simply $G_0 (i \omega , \vec{k} ) = [ i \omega -
\epsilon (\vec{k}) ]^{-1} $ and the interaction vertex is given by a
local repulsion, $U$, or some other short range interaction.  
The initial conditions are justified if $T_0$ is sufficiently
large.  We truncate the infinite system of equations by dropping all
vertex functions $\Gamma_T^{(n)}$ with $n>4$.  In the present
treatment we also neglect selfenergy corrections and the frequency
dependence of the vertex functions.  This restricts the scheme to
one--loop equations for the spin-rotation invariant four-point vertex
$\Gamma_T^{(4)}$.  Starting with weak to moderate interactions, we
follow the $T$-flow of $\Gamma_T^{(4)}$ as $T$ decreases.

The four-point vertex $\Gamma_T^{(4)}$ is determined by a coupling
function $V_{T} (\vec{k}_1, \vec{k}_2, $ $\vec{k}_3)\;$ (see
Refs.~\onlinecite{honerkamp,honerkampfm,sh}). 
The initial condition for pure on-site repulsion is 
$V_{T_0} (\vec{k}_1,\vec{k}_2,\vec{k}_3) \equiv U$.  
The numerical
implementation follows the work of Zanchi and Schulz~\cite{zanchi} and
was already explained in Refs.~\onlinecite{honerkamp} and \onlinecite{honerkampfm}.
We define elongated phase space patches around 
straight lines from the origin to the boundaries of the BZ. Next we 
approximate $V_{T} (\vec{k}_1,\vec{k}_2,\vec{k}_3)$ by a constant for all wave vectors 
in the same patch.  We calculate 
the RG flow for the discrete subset of interaction vertices with each
FS patch represented by a single wave vector. 
Most calculations were performed using 48 patches. We have checked the consistency 
of the discretization by comparing with up to 96 patches.   
Together with the flow of the interactions 
we calculate the flow of several static susceptibilities, as described
below.  This allows us to analyze 
which classes of coupling functions and which susceptibilities become
important at low $T$.  In many cases 
we observe a flow to strong coupling, i.e. at sufficiently low temperature some
components of the coupling function $V_{T}
(\vec{k}_1,\vec{k}_2,\vec{k}_3)$ become larger than the bandwidth.
The approximations mentioned above fail when the couplings become too
large.  Therefore we stop the flow when the largest coupling exceeds a
high value larger than the bandwidth, e.g. $V _{T, \mathrm{max}}=18t$.
This defines a {\em characteristic temperature} $T^*$ of the flow to strong coupling.
For superconducting and magnetic instabilities $T^*$ can be interpreted as estimate 
of the transition temperature if ordering is made possible, 
e.g. by including coupling in the third lattice direction. 

As in previous papers\cite{honerkampfm} we discuss the temperature flow at fixed chemical potential and not at fixed particle density. Since the relevant part of the flow only occurs at low temperatures we do not expect significant qualitative differences between the two possibilities. 

\section{Results at half filling and less} \label{boringFS}
\begin{figure}
\begin{center} 
\ifbuidln\includegraphics[width=.69\textwidth]{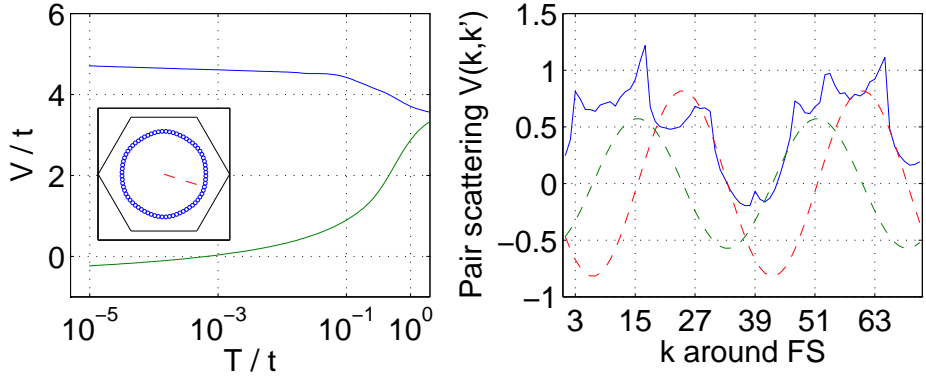}\fi
\end{center} 
\caption{Left: Flow of the most repulsive and most attractive coupling constant for $\mu=0.85t$ ($\langle n \rangle =1$/site). The inset shows the FS in the first BZ. Right: The solid line shows the pair scattering $V(\vec{k},-\vec{k} \to \vec{k}',-\vec{k}')$ at low $T$ with the patch index $k$ belonging to wavevector $\vec{k}$ around the 72 patches on the FS. The outgoing wavevector $\vec{k}'$ is fixed at point 3 (marked by the dashed line in the inset in the left plot). The dashed lines show the model pair scatterings $V(\theta_k, \theta_{k'=3})= -\sin 2 \theta_k  \sin 2\theta_3 $ and $V(\theta_k, \theta_{k'=3})=-\cos 2 \theta_k  \cos 2\theta_3 $ corresponding to the leading pairing susceptibilities with symmetries $sin 2 \theta_k$ and $\cos \theta_k$.}
\label{psplot85}
\end{figure}

\begin{figure}
\begin{center} 
\ifbuidln\includegraphics[width=.69\textwidth]{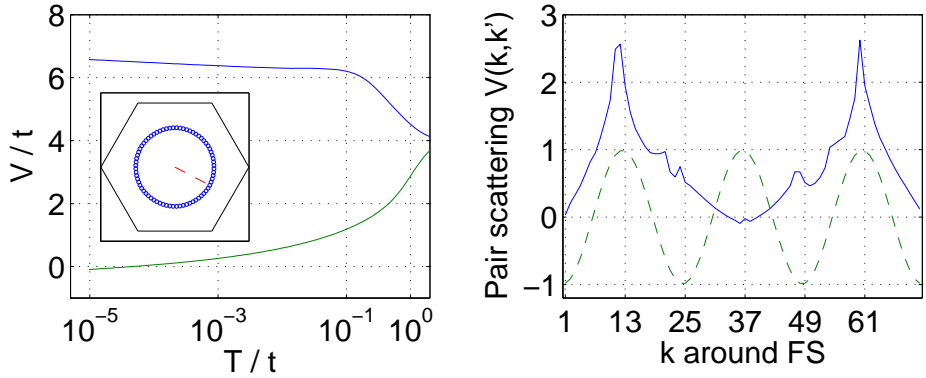}\fi
\end{center} 
\caption{Left: Flow of the most repulsive and most attractive coupling constant for $\mu=-0.7t$ ($\langle n \rangle \approx 0.65$/site). The inset shows the FS in the first BZ. Right: The solid line shows the pair scattering $V(\vec{k},-\vec{k} \to \vec{k}',-\vec{k}')$ at low $T$ with the patch index $k$ belonging to wavevector $\vec{k}$ around the 72 patches on the FS. The outgoing wavevector $\vec{k}'$ is fixed at point 1 (marked in the inset in the left plot). The dashed lines show the model pair scatterings $V(\theta_k, \theta_{k'=1})= -\sin 3 \theta_k  \sin 3\theta_1 $ corresponding to the leading pairing susceptibilities with symmetries $\sin 3\theta_k$.}
\label{psplot60}
\end{figure}

First let us consider band fillings where the Fermi surface is well inside the first BZ with particle densities around half filling and less. We start the temperature-flow RG at higher initial temperatures $\sim 4t$ with local on-site interactions $U=3.5t$ or $4t$ and follow the flow down to low scales. In Figs. \ref{psplot85} and \ref{psplot60} we display the behavior of the most repulsive and most attractive coupling constants. Although there is some flow, the coupling functions remain within the oder of the bandwidth $9t$, even if we go down to very low scales. Above $T \sim 10^{-5}t$ no tendency towards a flow to strong coupling, i.e. a divergence of the coupling constants, is observed.

Quite generally, for a non-nested Fermi surface away from van Hove singularities, superconductivity is the only possible instability at low scales in a weakly coupled system. Kohn and Luttinger\cite{kohn} gave an argument why superconductivity arises even in the presence of repulsive interactions. Although in our case we do not find a singularity over a wide scale range, it is still interesting to investigate which pair-scattering channel is most attractive. This channel can in principle become unstable at even lower temperatures. However in real systems these weak instabilities with non-$s$-wave pairing symmetries may be wiped out easily by impurity scattering.  

For the two different densities in Figs. \ref{psplot85} and \ref{psplot60} we plot the pair scattering $V(\vec{k},-\vec{k} \to \vec{k}',-\vec{k}')$ at low temperatures 
together with basis functions of the  most attractive pairing channels. We have calculated the temperature-flow of the pairing susceptibilities with angular dependence $\sin m \theta$ and $\cos m \theta$ for $m=1,2,3,\dots ,N$ around the FS.
For the half-filled case in Fig. \ref{psplot85} there are two harmonics that grow together more strongly than all other pairing channels. The dominant channels  are $d_{xy}=\sin 2 \theta$ and $d_{x^2-y^2} = \cos 2\theta$. These two $d$-wave functions form the basis of $\Gamma^+_6$, which is\cite{su} one of the two two-dimensional even-parity representations of the $D_{6h}$ symmetry of the hexagonal lattice\cite{fn2}. We expect that a possible superconducting pair amplitude will be proportional to a time-reversal symmetry-breaking superposition $d_{x^2-y^2}+id_{xy}$ of these two basis functions as this maximizes the condensation energy. Lee and Feng\cite{tklee} applied the Gutzwiller approximation scheme and variational Monte Carlo to the Heisenberg model on the triangular lattice.  Among the paired paramagnetic states they found the $d_{x^2-y^2}+id_{xy}$-wave state to have the lowest energy\cite{fn1}. Remarkably this is the pairing symmetry as the one suggested - albeit with extremely low $T_c$ - by the weak coupling RG at small to moderate $U/t$. The same type of superconducting pairing was also found near half band filling by various RVB theories\cite{baskaran,kumar,wang,ogata}.

We note that in this perturbative treatment of the on-site repulsion no precursors of a Mott transition at stronger $U$ are visible. This is different from the Hubbard model on the square lattice close to half filling, where for a certain parameter range the flow to strong coupling reveals tendencies towards incompressibility\cite{honerkamp}. 

Next let us consider the case $\langle n \rangle \approx 0.65$/site that might, as explained in Sec. I, correspond to the superconducting cobalt oxide compounds.  Here again no flow to strong coupling occurs down to very low temperatures $T \sim 10^{-5}t$ (see Fig. \ref{psplot60}). The dominant pair scattering is in an odd-parity triplet pairing channel with angle dependence $\sin 3 \theta$, corresponding to the $\Gamma^4_-$ representation of Ref. \onlinecite{su}. Comparing the basis function with the actual pair scattering rendered by the RG scheme, we notice that only two of the three repulsive portions of the $\sin 3 \theta$ around the FS are reflected in the pair scattering. Thus even though this representation is the most attractive pairing channel, is does not fully match the effective interactions which mix in other symmetries at this scale. The second strongest components are $d_{x^2-y^2}$ and $d_{xy}$, the first component is clearly visible for $\vec{k}'$ close to a maximum of $d_{x^2-y^2}$ in the right plot in Fig. \ref{psplot60}. Again, a possible $T_c$ in this case may be extremely low. 
  
We repeat that for our parameters and purely local interactions $0< U\le 4t$ we do not observe flows to strong coupling in the density range around and less than half filling. From the weak coupling perspective the reason simply is that the FS is relatively round and in the lack of nesting or van Hove singularities near the FS the one-loop corrections to the pair scattering do not generate any sufficiently attractive component in the pair scattering. Similar low energy scales for Kohn-Luttinger superconducting instabilities are found in the weakly repulsive Hubbard model on two-dimensional square lattice for a poorly nested Fermi surface away from the van Hove filling. 

We note that with our RG scheme which neglects the frequency dependence of the interaction vertices, it is difficult to investigate odd-frequency pairing. This unconventional type of pair formation\cite{balatsky} was suggested by calculations by Vojta and Dagotto\cite{vojta} for the Hubbard model on a triangular lattice for larger $U/t$. Although we cannot rule out this possibility for the time being, it is difficult to see how this type of instability should arise for the given rather generic situation unless drastic strong coupling effects are at work.

As has become clear, on-site repulsion alone does not give rise to superconductivity at reasonable energy scales for these band fillings. Therefore we extend our analysis and consider  exchange interactions between nearest neighbors,
\[ H_J = J \sum_{\langle ij \rangle} \vec{S}_i \cdot \vec{S}_j \]
with $\vec{S}_i = \frac{1}{2} \sum_{s,s'} \vec{\sigma}_{s,s'} c_{i,s}^\dagger c_{i,s'}$.
In principle these interactions will be generated by virtual hopping processes for local repulsion $U \gg t$. Since we treat all interactions perturbatively, this regime is not accessible with our RG method. We will therefore just assume that the initial action contains such an exchange term and analyze the consequences. 

\begin{figure}
\begin{center} 
\ifbuidln\includegraphics[width=.69\textwidth]{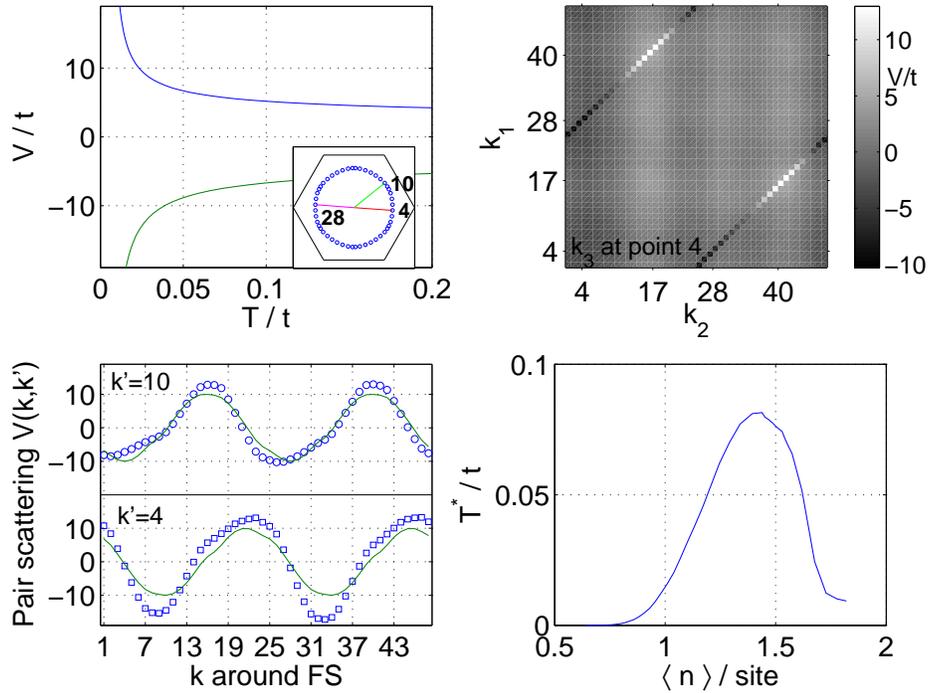}\fi
\end{center} 
\caption{Upper left plot: Flow of the most repulsive and most attractive coupling constant for $\mu=-0.85t$ ($\langle n \rangle \approx 1$/site), $U=0$ and $J=t$. The inset shows the FS in the first BZ. Upper right: 
Snapshot of the coupling function  $V_T (k_1,k_2,k_3)$ at low $T$. The first outgoing wavevector $\vec{k}_3$ is fixed at point 4
 and the wavevectors $\vec{k}_1$ and $\vec{k}_2$ vary around the Fermi surface. The diagonal features are the Cooper scattering processes with total incoming wavevector $\vec{k}_1+\vec{k}_2=0$. The scale-bar indicates the strength of $V_T (k_1,k_2,k_3)$.
Lower left:
The symbols show the pair scattering $V(\vec{k},-\vec{k} \to \vec{k}',-\vec{k}')$ at low $T$ with the patch index $k$ of wavevector $\vec{k}$ around the 48 patches on the FS. 
The outgoing wavevector $\vec{k}'$ is fixed at point 10 (circles in upper panel, see inset of upper left plot) and point 4 (squares in lower panel).
 The solid line shows the model pair scatterings $V(\theta_k, \theta_{k'})\propto  
- (\sin 2 \theta_k  \sin 2\theta_{k'} +\cos 2 \theta_k  \cos 2\theta_{k'}) $ corresponding to the leading pairing susceptibilities with symmetries $\sin 2\theta_k$ and $\cos2\theta_k.$
Lower right: Temperature scale for the $d_{x^2-y^2} +id_{xy}$ instability versus density for $U=0$ and $J=t>0$.}
\label{j1n}
\end{figure}

First let us consider the case $U=0$ and $J=t$ at half band filling. Now the interactions flow to strong coupling at $T\approx 0.02t$, and an analysis of the coupling functions and susceptibilities shows that the divergence is a very clean Cooper instability with again dominant $d_{x^2-y^2}+id_{xy}$ character. This is shown in Fig. \ref{j1n}. In the upper right plot we can see that only the coupling functions $V_T(\vec{k}_1,\vec{k}_2,\vec{k}_3)$ with total incoming wavevector $\vec{k}_1+\vec{k}_2=0$ grow rapidly towards low $T$. All other couplings remain smaller than the bandwidth. The pair scattering  $V(\vec{k},-\vec{k} \to \vec{k}',-\vec{k}')$ follows a $-( \cos 2 \theta_k \cos 2 \theta_{k'} + \sin 2 \theta_k \sin 2 \theta_{k'})$ form when the angles $\theta_k$, $\theta_{k'}$ are varied around the Fermi surface. This means that $d_{x^2-y^2}$ and $id_{xy}$ components, which belong to the same representation on the triangular lattice, get amplified in equivalent ways in a BCS gap equation with this pair scattering.

The same picture occurs for general band fillings. 
It turns out that at least for $J=t$, $U=0$, the $d_{x^2-y^2}+id_{xy}$-wave instability occurs over a wide density range from $\langle n \rangle =0.6$/site up to $\langle n \rangle =1.8$/site (see lower right plot of Fig. \ref{j1n}). 
The temperature scale for this instability is maximal near the van Hove filling. For $J=t$ it gets smaller than $10^{-5}t$ for densities less than $\langle n \rangle =0.6$/site, i.e. close to the density range $\langle n \rangle \approx 0.65$ relevant for the superconductivity in the cobalt oxides. Note however that this temperature scale depends exponentially on $J$ and the density of states close to the Fermi level.
We expect that an additional $U>0$ does not change the leading instability for these densities. We have checked this for $U=2t$, $J=0.5t$. Of course, our weak coupling analysis is not capable of describing a Mott state at half filling for large $U/t$.

It is interesting to observe the difference between the square lattice and the triangular lattice with nearest neighbor hoppings near half-filling. On the square lattice the flows to strong coupling and the suggested phases, mainly antiferromagnetic order and $d_{x^-y^2}$-wave superconductivity come out of the weak coupling RG very similarly for pure on-site $U$ or nearest neighbor exchange $J$.
On the triangular lattice the Fermi surface is not nested and no particle-hole symmetry exists. The flows for pure $U$ or pure $J$ interactions differ strongly in the  energy scales and, further away from half filling, also in the dominant pair scattering. 

Ferromagnetic exchange $J=-t$ induces a weak tendency towards $p$-wave superconductivity of $k_ +i k_y$-wave symmetry at 1/2 and 1/3 band filling. Around the van Hove filling we find a ferromagnetic instability.

\section{Around the van Hove filling}
For chemical potential $\mu= 2t$ or $3/4$ band filling, the non-interacting Fermi surface forms a hexagon. Its corners coincide with the van Hove singularities at the center points of the 6 flat sides of the BZ. Therefore we may expect some kind of infrared instability giving rise to a flow to strong coupling of a certain class of coupling constants. In addition to the diverging density of states, the flat sides  of the Fermi surface are perfectly nested, and this can cause a RG flow to strong coupling as well. 
The situation is reminiscent of the simple nearest-neighbor-hopping Fermi surface of the 2D square lattice at half filling, where two pairs of nested FS sides meet at the van Hove points at $(\pm \pi ,\pm \pi)$. However on the triangular lattice we expect that the nesting between the three pairs of opposite flat sides is somewhat less effective. The three nesting wavevectors (after addition of a reciprocal lattice vector) point to the three inequivalent centers of the BZ sides. Thus they do not correspond to the same wavevector as in the square lattice case. Therefore the low energy phase space weight gets distributed on three different spots in the BZ.
 
\begin{figure}
\begin{center} 
\ifbuidln\includegraphics[width=.99\textwidth]{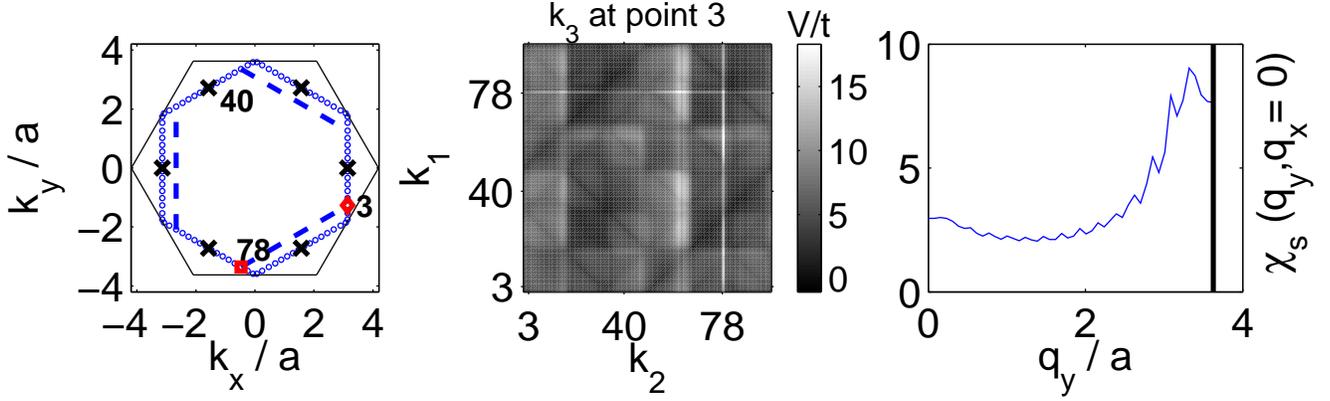}\fi
\end{center} 
\caption{Left: The first Brillouin zone of the triangular lattice and the Fermi surface (small circles for a $N=96$-patch calculation) at the van Hove filling $\mu=2t$, $\langle n \rangle =1.5$/site. The broken lines indicate the wavevector transfers that grow most strongly towards low temperatures. The crosses denote the 'hot' wavevectors $\vec{Q}_N$ near which the spin susceptibility grows most rapidly. Middle: Snapshot of the coupling function  $V_T ({k}_1,{k}_2,{k}_3)$ at low $T$. The first outgoing wavevector $\vec{k}_3$ is fixed near a van Hove point at point 3 (diamond in left plot), and the wavevectors $\vec{k}_1$ and $\vec{k}_2$ vary around the Fermi surface. The most repulsive scattering occurs for $k_2=78$ (square in left plot). Right: Spin susceptibility $\chi_s(\vec{q})$ at low $T$ along the line $q_x=0$, $q_y = 0 \dots 2\pi/3$. The enhancement is strongest near the nesting wavevector $\vec{Q}_N$ with $ q_y =2 \pi/\sqrt{3} \approx 3.63$ at the edge of the BZ.} 
\label{hex96_200}
\end{figure}

In Fig. \ref{hex96_200} we summarize the flow at the van Hove filling for $N=96$ patches. In the middle plot one observes several nesting features, i.e. bright, mainly vertical lines that correspond to fixed wave vectors transfer $\vec{k}_3 -\vec{k}_2$ between first outgoing and second incoming particles (the spin indices of first incoming and first outgoing have to be the same). The most divergent processes belong to umklapp scattering processes between the vicinity of (but not directly at) the van Hove singularities at the points where the FS touches the BZ boundary. The $\vec{k}_2$ and  $\vec{k}_3$ corresponding to the most divergent processes are connected by dashed lines in the left plot of Fig. \ref{hex96_200}. These processes drive the spin susceptibility, and correspondingly the spin susceptibility seems to diverge at these wavevectors. In the right plot of Fig. \ref{hex96_200} we show the $\vec{q}$ dependence of the static spin susceptibility along the line $q_x=0$, $q_y = 0 \dots 2 \pi/\sqrt{3}$. Compared to other regions in $\vec{q}$-space, it is enhanced along the whole line $q_x=0$, $q_y>0$ when the temperature is decreased, but shoots up most rapidly very close to $ q_y = 2 \pi/\sqrt{3} \approx 3.63$ at the edge of the BZ. This point and the other two symmetry-related points where the spin susceptibility diverges with equal strength are marked by crosses in  the left plot in Fig. \ref{hex96_200}. Since these wavevectors are equal to the three nesting vectors between the flat FS sides, we will denote them $\vec{Q}_N$. Due to the limited $\vec{q}$-space resolution of the $N$-patch discretization we are not able to decide clearly whether the peak occurs directly at or somewhat away from the BZ boundary. Going from $N=48$ to $N=96$ patches moves the peak a bit closer to the BZ edge, but for a clear answer not only the angular but also the radial dependence of the coupling constants must be considered. 
The susceptibility for ferromagnetism at $\vec{q}=0$ grows as well towards $T\to 0$, but the combination of large density of states and nesting makes the instability at $\vec{Q}_N$ far stronger.

A divergence of the spin susceptibility at a given wave vector signals strong ordering tendencies at the given wave vector, either in the ground state or, if possible, at finite temperatures.
In our case each of the $\vec{Q}_N$ would correspond to antiferromagnetic ordering along two directions of the triangular lattice and ferromagnetic ordering along the third lattice direction. From our calculation  we cannot tell whether the system will choose one of these possibilities, a certain superposition the three or whether ordering occurs at all. An analysis of the fluctuations around each of these orderings may give an answer to this question.

Next we decrease the particle density away from the van Hove filling. This reduces the density of states at the Fermi surface and cuts off the nesting. 
If the filling is sufficiently away from the van Hove filling we observe a saturation of the  spin susceptibility at all wave vectors before the interactions get larger than twice the bandwidth, which is our criterion for stopping the flow and beyond which the approximations may fail. For $U=4t$ this happens roughly for densities $\pm 0.03$/site and further away from the van Hove filling of 1.5 electrons/site. Obviously, like in the RG studies of the Hubbard model on the 2D square lattice\cite{zanchi,halboth,honerkampfm}, the precise boundaries of the spin density wave instability region depend somewhat on the scale where the susceptibilities are compared and on how far we trust the one-loop flow.

\begin{figure}
\begin{center} 
\ifbuidln\includegraphics[width=.69\textwidth]{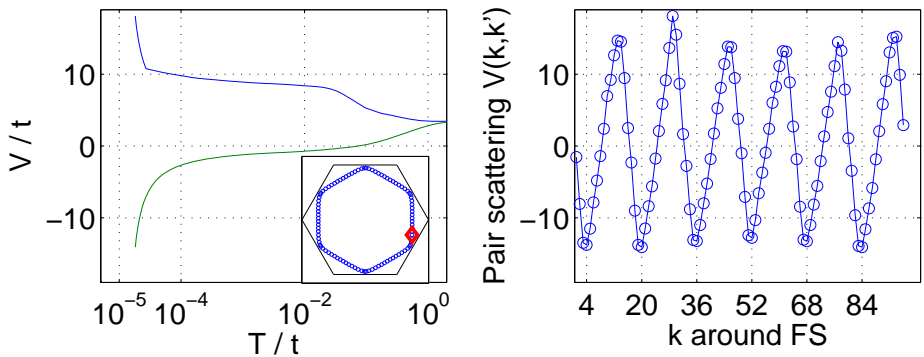}\fi
\end{center} 
\caption{Left: $T$-flow of most repulsive and most attractive coupling constant slightly away from the van Hove filling at $\mu= 1.94t$ and initial $U=3.5t$. The inset shows the 96 points on the FS used in the $N$-patch calculation. Right: pair scattering $V_T(\vec{k},-\vec{k} \to \vec{k}',-\vec{k}')$ at low $T$ with $\vec{k}'$ fixed at point 2 (marked with a diamond in the inset in the left plot) near but not at the van Hove points and $\vec{k}$ varying through the patches around the FS.}
\label{psplot189}
\end{figure}

\begin{figure}
\begin{center} 
\ifbuidln\includegraphics[width=.69\textwidth]{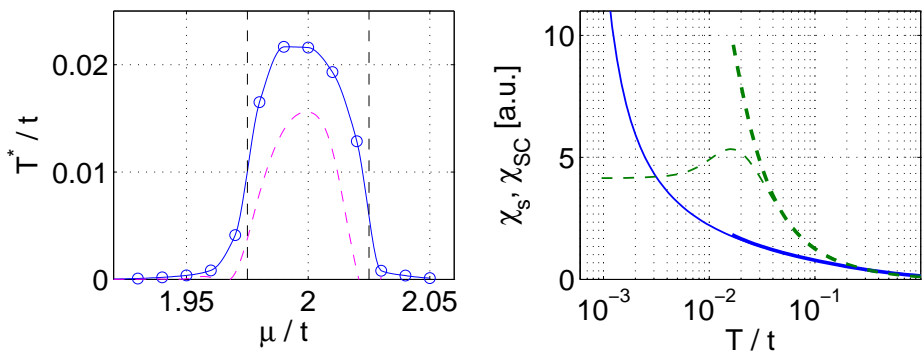}\fi
\end{center} 
\caption{Left: Temperature scale $T^*$ for the flow to strong coupling versus chemical potential $\mu$ for initial $U=4t$ (solid line with circles) or $U=3.5t$ (dashed line). Around the van Hove filling ($\mu=2t$) between the vertical broken lines (between densities 1.48 and 1.53 per site) the leading instability for $U=4t$ is in the spin channel near wavevectors $\vec{Q}_N$, outside these lines the dominant instability is singlet superconductivity. Left: Flow of the leading spin (near $\vec{Q}_N$, dashed lines) and superconducting susceptibilities (solid lines) for $\mu=1.98t$ (thick lines) and $\mu=1.96t$ (thin lines) for $U=4t$. }
\label{scmagmu}
\end{figure}

The pairing channels are not cut off, and if the flow in the particle-hole channel above the saturation is still strong enough, it generates a sizable attractive component in the pair scattering.  This component will flow in the particle-particle and eventually diverge in a Cooper instability at low temperatures. In Fig. \ref{psplot189} we plot the flow of the extremal coupling constants. We clearly observe the run-away flow of the coupling constants at $T\approx 2.5 \cdot 10^{-5} t$ that is solely due to the Cooper pair scattering processes. All other couplings, including those in the spin density wave channel, saturate at higher temperatures, as can be seen from the long plateau in the flow of the most repulsive coupling constant before the Cooper processes take off.
In the right plot of Fig. \ref{psplot189} we display the pair scattering around the Fermi surface. It has even parity in $\vec{k}$-space and roughly follows a $\sin 6 \theta$ behavior, corresponding to the $\Gamma^+_2$ representation of $D_{6h}$ according to Ref. \onlinecite{su}. The 12 nodes of the gap function are at the van Hove points and in the centers of the flat FS sides. The second strongest component in the pair scattering varies like $\sin 12 \theta$ around the Fermi surface.
The scale of the instability towards this unconventional $i$-wave state depends strongly on the vicinity to the van Hove filling. This is shown in Fig. \ref{scmagmu}. For larger on-site repulsion $U$, the density region with relevant values for $T^*$ becomes wider.

As we have seen at the end of the last section, an antiferromagnetic exchange interaction $J$ leads to a superconducting instability towards a time-reversal symmetry breaking state even at the van Hove density. Obviously there will be a change in the flow to strong coupling near the van Hove density when we change the initial interaction from $U>0$, $J=0$ to $U=0$, $J>0$. For $U=2t$ and $J=0.5t$, the instability is of $d_{x^2-y^2}+id_{xy}$-wave type. 
Here we do not attempt to determine the boundary between the spin density wave instability in the first case and the superconducting instability in the latter. Note however that this may be an interesting question for an approach that is capable of describing how the exchange term is generated by the Hubbard interaction.

\section{Conclusions}
We have analyzed the effective scattering processes at low temperatures of interacting electrons on the two-dimensional triangular lattice. We have used a 
$N$-patch temperature-flow renormalization group scheme in the one-loop 
approximation that has been found useful\cite{zanchi,halboth,honerkampfm} 
in the analysis of the weakly coupled Hubbard  model on the 2D square lattice. 
Starting with a weak to moderate on-site repulsion $U\sim 3 \dots 4 t$ at high temperatures, we have investigated possible magnetic and superconducting instabilities when the temperature is lowered. For most fillings and purely local initial interactions, especially for densities less than $\approx$1.4 electrons per lattice site (for hopping parameter $t>0$), the interactions remain within the order of the bandwidth. Down to very low temperatures $T \sim 10^{-5}t$ no flows to strong coupling are observed.  Replacing the pure on-site Hubbard interaction by an antiferromagnetic exchange interaction $J$, the RG flow leads to  a superconducting instability towards a time-reversal symmetry breaking $d_{x^2-y^2}+id_{xy}$-wave state at sensible scales for a wide density range between $\langle n \rangle > 0.6$/site and $\langle n \rangle > 1.8$/site.
This state has the same pairing symmetry as the superconducting RVB states found in recent theories for strong on-site repulsion\cite{baskaran,kumar,wang,ogata}. It may also be obtained by ordinary mean-field theories that do not invoke spin-charge separation. 

Near the van Hove density $\langle n \rangle = 1.5$/per site, we observe a flow to strong coupling also for purely local Hubbard interactions. The rapid growth of the interactions at low temperatures is driven by umklapp processes between the van Hove regions and  nesting between the flat parts of the (nearly) hexagon-shaped Fermi surface. In the immediate  vicinity of the van Hove filling the spin susceptibility near the wavevector $\vec{q}=(0, 2\pi/3)$ and two symmetry-related  wavevectors diverges most strongly, signalling magnetic ordering tendencies at these wavevectors. It may be interesting to investigate whether the systems tends to order (if we allow coupling between layers) by choosing one wavevector or by forming a superposition of the three.
Slightly away from the van Hove filling (outside the range  $1.5\pm0.03$/site for $U=4t$), the flow in the spin density wave channel is cut off, and the dominant instability occurs in a singlet pairing channel with symmetry $\Delta (\theta) \propto \sin 6 \theta$, having 12 nodes and 12 extrema when the angle $\theta$ moves around the  Fermi surface. This superconducting instability occurs on both sides of the van Hove filling and the characteristic temperature $T^*$ falls of rapidly with increasing distance to the van Hove density. Since the superconducting state has nodes  in the vicinity  of the van Hove points, it may be susceptible to further instabilities at lower temperatures which try to remove the large density of states at zero energy.

Turning to the reported superconductivity in the layered cobalt oxides, we have to keep in mind that it is far from settled that these materials can be described by a simple one-band Hubbard model on the triangular lattice. Neither is it clear that they can be treated using weak-coupling techniques. In fact estimates\cite{singh} of the bandwidth and Hubbard $U$ for NaCo$_2$O$_4$ seem to show that the material is strongly correlated. 
Nevertheless it may be helpful to know the weak coupling picture.
Band structure calculations for NaCo$_2$O$_4$ by Singh \cite{singh} indicate that the main Fermi surface is hole-like and centered around the $(0,0)$ point. Angular resolved photoemission data support this picture\cite{valla}.  Translated into the language of a tight-binding model this case corresponds to a negative hopping parameter $t<0$ for electrons. The superconducting samples may be at $\approx$1.35 electrons/Co site, or, with our sign convention $t>0$ that interchanges electrons and holes $\langle n \rangle =0.65$/site. 
This case is investigated in Section \ref{boringFS}. For a purely local Hubbard repulsion $U>0$ without additional exchange interactions no sizable superconducting tendencies can be detected with our weak coupling technique. 
Antiferromagnetic exchange interactions $J>0$ generate a superconducting instability towards a $d_{x^2-y^2}+id_{xy}$-wave state over a wide density range. 
In this respect, weak coupling RG and RVB theories for the strongly correlated $t$-$J$ model\cite{baskaran,kumar,wang,ogata} give similar predictions at least as far as the symmetry of the pairing is concerned. 
Thus the time-reversal symmetry breaking $d+id'$ state may be the primary candidate for experimental checks. 
This state has a fully gapped single-particle excitation spectrum. The phase structure of the order parameter may give rise to observable effects\cite{sigrist} such as subgap Andreev bound states at surfaces similar to those in $p$-wave superconductors\cite{honsig,kashiwaya}.

The main purpose of the present work was to investigate possible superconducting states on the triangular lattice at weak to moderate interactions. Further work is needed in order to clarify the relevance of the weak coupling physics to real materials like the cobalt oxides.
\\*[2mm]
I thank P.A. Lee, V. Liu, O. Motrunich, T.M. Rice, T. Senthil, and A. Vishwanath for helpful discussions. The Deutsche Forschungsgemeinschaft (DFG) is acknowledged for financial support.


\begin{thebibliography}{99}
\bibitem{fn2} If we restrict the study to purely 2D systems, the representations that are odd with respect to reflections at the $x$$y$-plane have to be discarded.
\bibitem{fn1}
Lee and Feng\cite{tklee} normalize their basis functions differently. In their notation, $d_{x^2-y^2}+i\sqrt{3}d_{xy}$ is the state with the lowest energy.
\end{thebibliography}
\end{document}